\documentstyle[floats,aps,prl,twocolumn,epsf]{revtex}

\begin{document}                % INITIALIZE - DONT CHANGE
\draft
\preprint{}
\twocolumn[\hsize\textwidth\columnwidth\hsize\csname
@twocolumnfalse\endcsname
\title{Photoemission Spectral Weight Transfer and Mass Renormalization
  in the Fermi-Liquid System La$_{1-x}$Sr$_x$TiO$_{3+y/2}$}
\author{T. Yoshida$^1$, A. Ino$^1$, T.  Mizokawa$^2$, A. Fujimori$^{1,2}$,
Y.
  Taguchi$^3$, T.  Katsufuji$^3$, Y. Tokura$^3$}
\address{$^1$Department of Physics, University of Tokyo, Bunkyo-ku,
  Tokyo 113-0033, Japan}
\address{$^2$Department of Complexity Science
  and Engineering, University of Tokyo, Bunkyo-ku, Tokyo 113-0033,
  Japan}
\address{$^3$Department of Applied Physics, University of
  Tokyo, Bunkyo-ku, Tokyo 113-0033, Japan}
\date{\today}
\maketitle

\begin{abstract}                % DON'T CHANGE THIS LINE
  We have performed a photoemission study of
  La$_{1-x}$Sr$_x$TiO$_{3+y/2}$ near the filling-control
  metal-insulator transition (MIT) as a function of hole doping.  
  Mass renormalization deduced from the spectral weight and the
  width of the quasi-particle band around the chemical potential $\mu$ 
  is compared with that deduced from
  the electronic specific heat.  The result implies that, near the
  MIT, band narrowing occurs strongly in the vicinity of $\mu$.
  Spectral weight transfer occurs from the coherent to the incoherent
  parts upon antiferromagnetic ordering, which we associate with the
  partial gap opening at $\mu$.
\end{abstract}

]

\narrowtext
%\section{INTRODUCTION}               % Introduction goes below.
The critical behavior of the electronic structure in the vicinity of
the filling-control metal-insulator transition (MIT) has attracted
much attention because of their fundamental importance \cite{intro}
and their relevance to high-$T_c$ superconductivity in layered
cuprates. La$_{1-x}$Sr$_x$TiO$_3$ is a suitable system for
investigating the filling-control MIT; their thermodynamic and
transport properties have been systematically studied
\cite{tokura1,kumagai}.  The system evolves from a Mott-Hubbard-type
insulator having the $d^1$ configuration (LaTiO$_3$) to a band
insulator having the $d^0$ configuration (SrTiO$_3$) with a wide range
of the paramagnetic metallic (PM) phase in-between. Here, LaTiO$_3$ is
antiferromagnetic below the N\'{e}el temperature $T_N\sim$140 K
\cite{tokura2}. Holes can be doped into LaTiO$_3$ by Sr substitution
for La and/or by excess oxygens as in La$_{1-x}$Sr$_x$TiO$_{3+y/2}$
(LSTO), where $\delta = x+y$ is the hole concentration. The electrical
resistivity $\rho$ in the PM phase shows a $T^2$ dependence,
characteristic of an interacting Fermi liquid
\cite{tokura1,katsufuji}. With decreasing $\delta$ in the PM phase
($0.08 < \delta < 1$), the electronic specific heat coefficient
$\gamma$, which is proportional to the conduction electron effective
mass $m^*$, is enhanced towards the antiferromagnetic (AF) phase
boundary ($\delta$= 0.08) and then decreases in the
antiferromagnetic metallic (AFM) phase ($0.05< \delta <0.08$) and the
antiferromagnetic insulating (AFI) phase ($\delta < 0.05$)
\cite{kumagai,taguchi}.  The enhancement of $\gamma$ is expected for
correlated metals near a Mott transition, and has been analyzed within
the Fermi-liquid framework \cite{tokura1,kumagai}.

Photoemission studies of the metallic bandwidth-control system
Ca$_x$Sr$_{1-x}$VO$_3$ with the fixed band filling ($d^1$) \cite{huga}
have indicated that the V 3$d$ band is split into the quasi-particle
band and the remnant of the Hubbard bands (the coherent and incoherent
parts, respectively, of the spectral function) and that spectral
weight transfer occurs between them as a function of $U/W$, where $U$
is the one-site Coulomb repulsion and $W$ is the one-electron band
width, consistent with the prediction of dynamical mean-field theory
(DMFT) \cite{zhang}. However, the effective mass obtained from the
electronic specific heat coefficient $\gamma$ and the magnetic
susceptibility $\chi$ shows only a very weak enhancement with
increasing $U/W$ \cite{huga2}. The Ti 3$d$ band in the photoemission
spectra of the filling-control systems LSTO \cite{fujimori} and
Y$_{1-x}$Ca$_x$TiO$_3$ \cite{morikawa} also shows the same type of
splitting into the coherent and incoherent parts, as has been
predicted by DMFT \cite{kotliar}.  In the present work, we have made a
detailed photoemission study of LSTO ranging from the PM to AFM to AFI
phases in order to clarify the electronic structure near the
filling-control MIT.  In particular, on the PM side of the PM-AFM
phase transition, we discuss its effective mass renormalization based
on the spectroscopic data and compare it with that deduced from the
thermodynamic measurements.  Since LSTO can be viewed as
a normal Fermi liquid in the PM regime, it is interesting to compare
its spectroscopic properties with those of the high-$T_c$ cuprates,
which show a remarkable deviation from the normal Fermi liquid in the
underdoped regime. Therefore, we compare our results on LSTO with
previous photoemission results on La$_{2-x}$Sr$_x$CuO$_4$ (LSCO)
which show pseudogap behavior in the underdoped regime \cite{ino}.
The filling-control MIT's in LSTO and LSCO may be considered as
representative of the two types of Mott transitions caused,
respectively, by a divergent effective mass $m^*$ and a vanishing
carrier number $n$ \cite{Imada}.

\begin{figure*}[!t]
\epsfxsize=175mm\centerline{\epsfbox{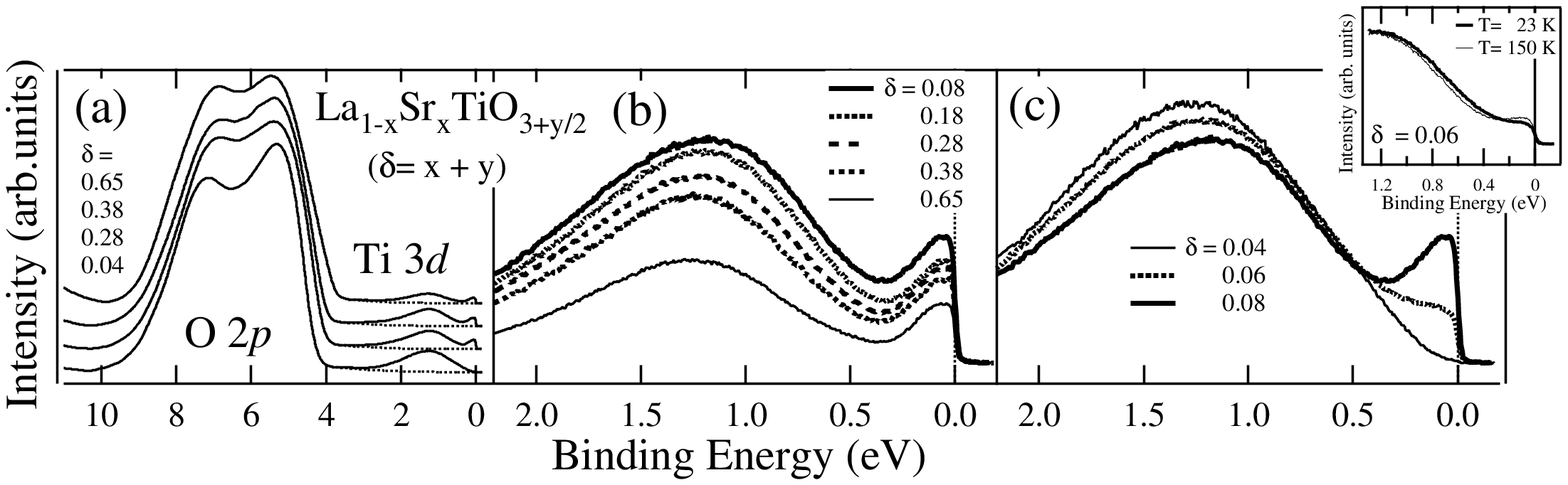}}\vspace{0.5pc}
    \caption{Valence-band photoemission spectra of
La$_{1-x}$Sr$_x$TiO$_{3+y/2}$ (a). The dotted curves are assumed backgrounds.
Photoemission spectra
      of La$_{1-x}$Sr$_x$TiO$_{3+y/2}$ in the Ti 3$d$ band region in
      the high doping regime (b) and the low doping regime (c). The
      inset shows the spectra of the AFM sample ($\delta$=0.06) below
      and above $T_N=$ 112 K.}
\label{Spectra}
\end{figure*}

%experiment
Samples of La$_{1-x}$Sr$_x$TiO$_{3+y/2}$ and LaTiO$_{3+y/2}$ were
melt grown by the floating-zone method in a reducing or inert
atmosphere \cite{fujishima}. LaTiO$_{3+\delta/2}$ covers $\delta $=
0.04, 0.06, and 0.08 and La$_{1-x}$Sr$_x$TiO$_{3+y/2}$ covers $\delta$=
0.18, 0.28, 0.38, and 0.65. The $\delta$ values were determined by
thermogravimetric analysis. Ultraviolet photoemission
spectroscopy measurements were performed using a VSW
hemispherical analyzer and a helium discharge lamp ($h\nu$= 21.2 eV).
The total energy resolution was $\sim$30 meV. The base pressure in
the spectrometer was in
the $10^{-10}$ Torr range. Clean surfaces were obtained by {\it in situ}
scraping with a diamond file.  The electron chemical potential $\mu$,
namely, the Fermi level $E_F$ was determined from
the spectra of Au evaporated on each
sample.  The measurements
were made at 20 K unless otherwise stated.

%\section{Valence Band spectra}
The valence-band photoemission spectra of LSTO show structures due to
the O 2$p$ and Ti 3$d$ bands as shown in Fig.~\ref{Spectra}(a).  In order
to extract the Ti 3$d$ band, we have assumed the tail of the O 2$p$
band as shown by dotted curves in the figure.  Figure~\ref{Spectra}(b)(c)
shows the Ti 3$d$-band spectra thus deduced, whose integrated
intensity has been normalized to the band filling $1- \delta$.  This
normalization and the normalization to the integrated intensity of the
O 2$p$ band gave nearly identical results.  The peak within $\sim$0.3
eV of $\mu$ is the coherent part due to the quasi-particle excitation
and corresponds to the renormalized Ti $3d$ band.  The broad peak
$\sim$1.1 eV below $\mu$ is the incoherent part, reminiscent of the
lower Hubbard band corresponding to the $d^1\to d^0$ spectral weight
of the insulating LaTiO$_3$ \cite{zhang}.

One can see that in going from $\delta=0.65$ to 0.08 within the PM
phase, both the coherent and incoherent parts become stronger
[Fig .~\ref{Spectra}(b)], and in going from $\delta=0.08$ to 0.06 in
the AFM phase, the coherent part becomes weaker
[Fig.~\ref{Spectra}(c)].  We consider that in the AFM phase a gap
partially opens on the Fermi surface due to the AF ordering.  As shown
in the inset of Fig.~\ref{Spectra}(c), we observed for $\delta=0.06$
that the spectral weight of the coherent part increases in going from
below $T_{N}$ (= 112 K) to above it.  This fact implies that the
suppression of the DOS of the coherent part is indeed due to the AF
ordering and would be related to the increase of the resistivity of
the $\delta=0.06$ sample below $\sim$100 K \cite{taguchi}.  In the
spectrum of the AFI ($\delta=0.04$) sample, the spectral weight of the
coherent part vanishes.  Note that the spectral weight transfer from
the coherent to the incoherent parts as shown in
Fig.~\ref{Spectra}(c) involves energies as high as $\sim1$ eV and
cannot be explained by the ordinary spin-density-wave theory. Electron
correlation effects would have to be invoked to explain such
behavior across the PM-AFM-AFI transitions.

\begin{figure}[!t]
\centerline{\epsfxsize=80mm \epsfbox{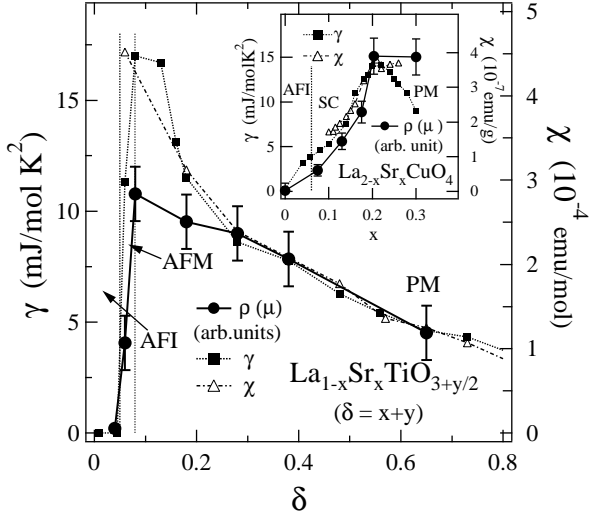}} \vspace{0.5pc}
    \caption{Comparison of the DOS at $\mu$, $\rho(\mu)$, with
      the electronic specific heat coefficient
      $\gamma$ and the Pauli-paramagnetic susceptibility
      $\chi$\protect\cite{tokura1,kumagai,taguchi}.
     The inset shows $\rho(\mu)$, $\gamma$ and $\chi$ of
     LSCO\protect\cite{ino,momono,nakano}.}
\label{gamma}
\end{figure}

%comparison with gamma
In Fig.~\ref{gamma}, the doping dependence of the spectral density of
states (DOS) at $\mu$, $\rho(\mu)$, is compared with
the electronic specific heat coefficients $\gamma$ and the Pauli
paramagnetic susceptibility $\chi$\cite{tokura1,kumagai,taguchi}. One
can see from the figure that $\rho(\mu)$, $\gamma$, and
$\chi$ show qualitatively similar doping dependencies; they increase
with decreasing $\delta$ in the PM phase until the PM-AFM
phase boundary ($\delta$= 0.08) is reached. However, a closer inspection
reveals that while they all increase in the same way in the high
doping regime ($\delta > 0.3$), $\gamma$ and $\chi$ show faster
increases than $\rho(\mu)$ in the low doping regime ($\delta<0.3$) with
decreasing $\delta$.

\begin{figure}[!t]
 \centerline{\epsfxsize=73mm \epsfbox{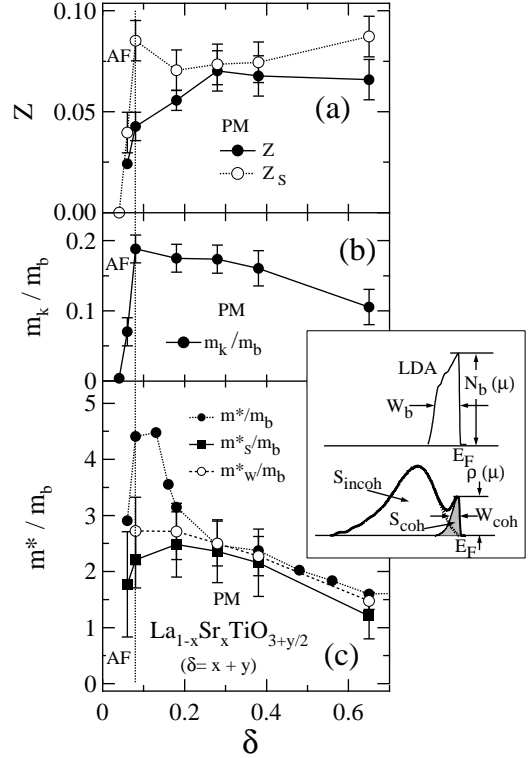}} \vspace{0.5pc}
\caption{Renormalization factor $Z$ (a),  the
$k$-mass $m_k$ (b), and the thermodynamic
effective mass $m^*$ (c) derived from various experiments.
The inset shows how the lineshape of the Ti $3d$ band has been
analyzed when deriving $m_S^*/m_b$ [$=(m_k/m_b)(1/Z_S) =
(N_b(\mu)/\rho(\mu))(1+S_{incoh}/S_{coh})$] and
$m_W^*/m_b$ ($=W_b/W_{coh}$).}
    \label{zmass}
\end{figure}

%formulation
In order to obtain further information about the mass renormalization
of the conduction electrons in LSTO,
we have evaluated the quasi-particle weight $Z$, the $k$-mass
$m_k$, and the thermodynamic effective
mass $m^*$ \cite{huga,morikawa,greeff} using the
relationship $Z=\rho(\mu)/N^*(\mu)$,
$m_k / m_b = \rho(\mu) / N_b(\mu)$, and $m^* / m_b =
(1/Z)( m_k / m_b)$. Here, $N^*(\mu)$ is the density of quasi-particles
deduced from $\gamma = (\pi^2/3 k_B^2)N^*(\mu)$,
$m_b$ is the band mass, and $N_b(\mu)$ is the
band DOS at $\mu$ derived from the PM band structure
calculated using the local-density approximation (LDA)
\cite{takegahara}.
$Z$ and $m_k/m_b$ are
given in terms of the self-energy $\Sigma({\bf k},\omega)$
by $Z= (1-\partial{\rm Re}\Sigma({\bf k},\omega) /
\partial\omega\vert_{\omega=\mu,k=k_F})^{-1}$
and $m_k/m_b =(\partial \varepsilon_k /\partial k)
/[\partial \varepsilon_k / \partial k +
\partial {\rm Re}\Sigma ({\bf k},\omega) / \partial
k]\vert_{\omega=\mu,k=k_F}$.
Alternatively, we have evaluated $Z$ by assuming that
the spectral weight of the coherent part $S_{coh}
\propto Z$ and that of the incoherent part $S_{incoh}
\propto 1-Z$, namely, by assuming that the mass
renormalization is uniform throughout the quasi-particle band.
Here, the Ti $3d$ band has been divided into the
coherent and incoherent parts as shown in the inset of
Fig.~\ref{zmass}. We denote this $Z$ by $Z_S$.

%renormalization factor
The doping dependence of $Z$ and $Z_S$ thus obtained is shown in
Fig.~\ref{zmass}(a).  For $\delta>0.2$, both $Z$ and $Z_S$ are small
($\sim$0.06), indicating strong correlation effects.  $Z$ further
decreases with decreasing $\delta$ in the low doping regime ($\delta
< 0.2$), indicating the increase of electron correlation effects.
Since the decrease of $Z$ is pronounced near the PM-AFM
phase boundary, the effect of AF fluctuations may be
responsible for it. On the other hand, $Z_S$ does not
decrease near the PM-AFM transition.
The difference between $Z$ and $Z_S$ in the low doping
regime means that the assumption of the uniform mass renormalization
throughout the quasi-particle band breaks down in this $\delta$ regime.
From this, we may conclude that band narrowing is particularly
enhanced in the vicinity of $\mu$ near the
PM-AFM boundary.  Very recently, an angle-resolved photoemission study
of Mo(110) surface states has shown that the
band dispersion becomes less steep within $\sim$50 meV of $\mu$,
indicating a strong energy dependence of
$\partial{\rm Re}\Sigma({\bf k},\omega) / \partial\omega$
in the vicinity of $\mu$, which has been attributed
to electron-phonon interaction \cite{johnson}.
In Fig.~\ref{zmass}(b), the $k$-mass
$m_k/m_b$ shows a weak increase with decreasing $\delta$ in the PM phase.
This is contrasted with the bandwidth-control MIT in Ca$_{1-x}$Sr$_x$VO$_3$,
where $m_k$ rapidly decreases with increasing $U/W$ \cite{huga}.

%Effective mass
We then evaluate the mass enhancement factor $m^*/m_b$ from the
spectroscopic data using the two methods: $m_S^* / m_b = (m_k
/m_b)(1/Z_S)$ and $m_W^* / m_b = W_b / W_{coh}$
\cite{huga,morikawa,greeff}, where $W_b$ is the bare band width from
the LDA band-structure calculation\cite{takegahara} and $W_{coh}$ is
the quasi-particle band width observed by the photoemission experiment
(see the inset of Fig~\ref{zmass}). In Fig.~\ref{zmass}(c), the doping
dependence of $m_S^*/m_b$ and $m_W^*/m_b$ are compared with the
thermodynamic effective mass $m^{*}/m_{b}=N^{*}(\mu)/N_{b}(\mu)$
deduced from $\gamma$. The doping dependence of $m^{*}/m_b$ is weak in
the high doping regime ($\delta > 0.3$) while it becomes stronger in
the low doping regime ($\delta < 0.3$). In the high doping regime,
where $Z_S\simeq Z$, both $m_S^*/m_b$ and $m_W^*/m_b$ agree with
$m^*/m_b$, indicating that the quasi-particle band is uniformly
narrowed.  Near the PM-AFM phase boundary, on the other hand,
$m_S^*/m_b$ and $m_W^*/m_b$ are not so enhanced as $m^*/m_b$. This
again indicates that the band narrowing does not occur uniformly by
the factor $m_S^*/m_b$ or $m_W^*/m_b$ in the quasi-particle band,
but more strongly in the vicinity of $\mu$.

While the overall behavior of the spectral lineshapes and the
spectroscopic and thermodynamic mass renormalization in LSTO is
consistent with the prediction of DMFT \cite{kotliar}, some
discrepancies are noted: DMFT predicts a rather uniform band narrowing
and a constant $m_k/m_b$ ($=1$), leading to $Z \simeq Z_S$ and $m^*
\simeq m^*_S \simeq m^*_W$ in the entire doping range. The 
discrepancies between $m*$, $m_S^*$ and $m_W^*$
described above, particularly in the low doping regime,
indicate that AF fluctuations, which are not included in DMFT, may
play some role near the MIT.

%surface
Now, we consider the effect of surface states
on the above analyses for LSTO, because there is a possibility that
the incoherent part includes contributions from the surface states as
in the case of the perovskite-type La$_{1-x}$Ca$_x$VO$_3$ (LCVO)
\cite{maiti}.  Surface contributions in LSTO should be smaller than in
LCVO, because the O 1$\it s$ core-level spectra of LSTO show a
single peak for all the samples, in contrast to LCVO which shows
multiple peak structures owing to possible charge disproportionation
into V$^{3+}$ and V$^{5+}$ on the surface.  Nevertheless, we have
estimated to what extent the present analyses are altered if
surface contributions exist.  We assume that the surface-derived
spectral weight in the Ti $3d$ band region is $\alpha S (\alpha<1)$
compared to the bulk contribution $S\equiv S_{coh}+S_{incoh}$, and that
the surface signals contribute only to the incoherent part as in the
case of LCVO \cite{maiti}.  Under this assumption, $\rho(\mu)$ and
$S_{coh}$ should be multiplied by $1/(1-\alpha)$ and hence $Z$, $Z_S$
and $m_k/m_b$ by $1/(1-\alpha)$ as a correction
for the surface contribution.  Hence, the doping dependence of these
quantities would not change although their absolute values are
multiplied by $1/(1-\alpha)$.  Furthermore, $m^*_S/m_b
=(m_k/m_b)(1/Z)$ would not change by this correction because the
factor $1/(1-\alpha)$ cancels out.  $m_W^*/m_b$ does not change, too, since
it has been estimated by using only the band widths. Therefore we can
say that if the surface contribution $\alpha$ is constant in the
doping range considered here, the doping dependence of $Z$, $Z_S$, and
$m_k/m_b$ evaluated above remains valid, and the mass enhancement
factor $m^*_S/m_b$ or $m_W^*/m_b$ remains unchanged after the
correction.

%comparison with LSCO
It is interesting to compare the doping dependence of the mass
renormalization in LSTO and that in LSCO in order to clarify what is
common and what is different between the conventional Fermi liquid and
the non-Fermi liquid of the underdoped cuprates.  In the overdoped
regime of LSCO, $\gamma$\cite{momono} and $\chi$ \cite{nakano}
increase with decreasing doping for $\delta >0.2$, as shown in the
inset of Fig.~\ref{gamma}. This enhancement is similar to that in
LSTO.  However, for $0.1< \delta <0.2$, where LSCO is still in the PM
(or superconducting) phase, $\gamma$, $\chi$, and $\rho(\mu)$ decrease
with decreasing $\delta$ because of the pseudogap opening \cite{ino}.
Such pseudogap behavior has not been identified in the PM state of
LSTO, although a very weak pseudogap has been postulated from magnetic
susceptibility measurements at high temperatures ($>$300 K)
\cite{ono1}. The fact that the suppression of the density of
quasi-particles in the PM phase of LSCO is comparable to that in the
AFM state of LSTO may indicate that AF spin correlations in LSCO are
strong enough to reduce the density of quasi-particles at $\mu$.  LSTO
can therefore be considered as a normal Fermi liquid with negligible
AF fluctuations except for near the PM-AFM phase boundary, while for
$\delta<0.08$ the static AF order suppresses the DOS as described
above. In spite of the contrasting behavior of LSTO and LSCO, $Z$
decreases with decreasing $\delta$ towards the PM-AFM phase boundary
in both systems. It therefore appears that the decrease of $Z$, namely
the loss of coherent spectral weight at and around $\mu$, is a common
feature of the two types of Mott transitions.

%\section{CONCLUSION}
In conclusion, in the valence-band photoemission spectra of LSTO, the
spectral DOS at $\mu$, $\rho(\mu)$, is enhanced towards the PM-AFM
phase boundary in the PM phase, but is dramatically reduced in the AFM
phase. In going from the PM to the AFM phases, spectral weight
transfer takes place from the coherent part to the incoherent part.
From the analyses of mass renormalization, it appears that near the
PM-AFM transition, the band narrowing does not occur uniformly
throughout the quasi-particle band, but preferentially in the vicinity
of $\mu$.  The doping dependencies of $\rho(\mu)$, $m^*/m_b$,
$\gamma$, and $\chi$ of LSTO are contrasted with those of LSCO.
Although LSTO shows a systematic enhancement of quasi-particle density
with decreasing $\delta$ in the PM phase, LSCO shows a suppression
with decreasing $\delta$ in the PM (superconducting) phase at
$\delta<0.2$.  These facts imply the importance of AF spin correlation
in LSCO in this doping range.

%\section{ACKNOLEDGEMENT}
We would like to thank Y. Takada and M. Imada for useful discussions and
K. Kobayashi and T. Susaki for help in the measurements.
This work is supported by a Special Coordination Fund
from the Science and Technology
Agency and the New Energy and Industrial Technology
Development Organization (NEDO).

\end{document}